\documentclass{article}
\usepackage[preprint]{spconfa4}
\usepackage{spconfa4,amsmath,graphicx}
\usepackage{booktabs}
\usepackage{hyperref}
\toappear{\begin{minipage}{\textwidth}\footnotesize \copyright2024 IEEE.  Personal use of this material is permitted.  Permission from IEEE must be obtained for all other uses, in any current or future media, including reprinting/republishing this material for advertising or promotional purposes, creating new collective works, for resale or redistribution to servers or lists, or reuse of any copyrighted component of this work in other works.\end{minipage}}

\title{DSP-informed bandwidth extension using locally-conditioned excitation and linear time-varying filter subnetworks}

\name{Shahan Nercessian, Alexey Lukin, and Johannes Imort}
\address{Native Instruments}
\begin{document}
%\ninept
%
\maketitle
\begin{abstract}
In this paper, we propose a dual-stage architecture for bandwidth extension (BWE) increasing the effective sampling rate of speech signals from 8~kHz to 48~kHz.  Unlike existing end-to-end deep learning models, our proposed method explicitly models BWE using excitation and linear time-varying (LTV) filter stages.   The excitation stage broadens the spectrum of the input, while the filtering stage properly shapes it based on outputs from an acoustic feature predictor.  To this end, an acoustic feature loss term can implicitly promote the excitation subnetwork to produce white spectra in the upper frequency band to be synthesized. Experimental results demonstrate that the added inductive bias provided by our approach can improve upon BWE results using the generators from both SEANet or HiFi-GAN as exciters, and that our means of adapting processing with acoustic feature predictions is more effective than that used in HiFi-GAN-2. Secondary contributions include extensions of the SEANet model to accommodate local conditioning information, as well as the application of HiFi-GAN-2 for the BWE problem.

\end{abstract}

\begin{figure*}[t]
  \begin{minipage}{.33\textwidth}
	\centering
  \centerline{\includegraphics[width=0.7\columnwidth]{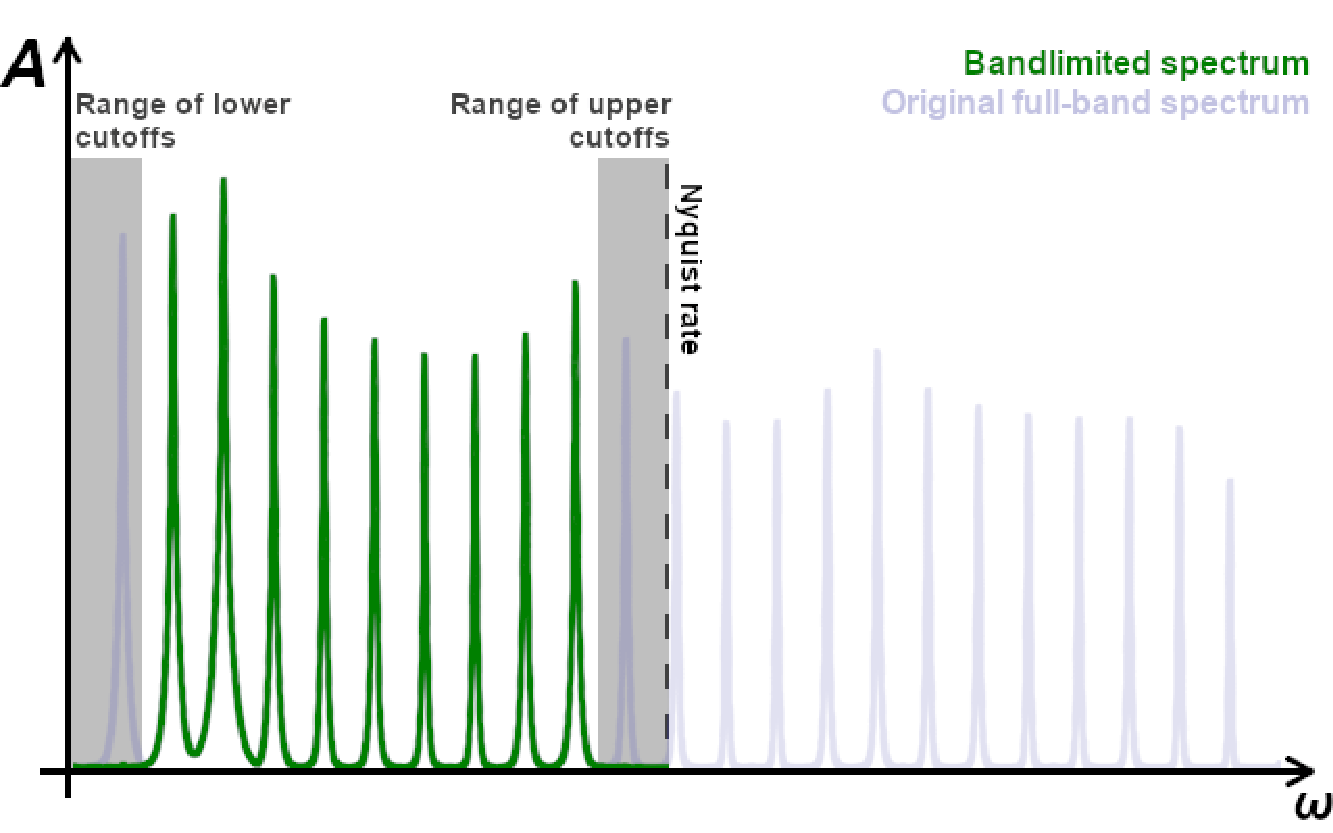}}
	\vspace{-0.8em}
  {(a)}
  \end{minipage}
  \begin{minipage}{.33\textwidth}
	\centering
  \centerline{\includegraphics[width=0.7\columnwidth]{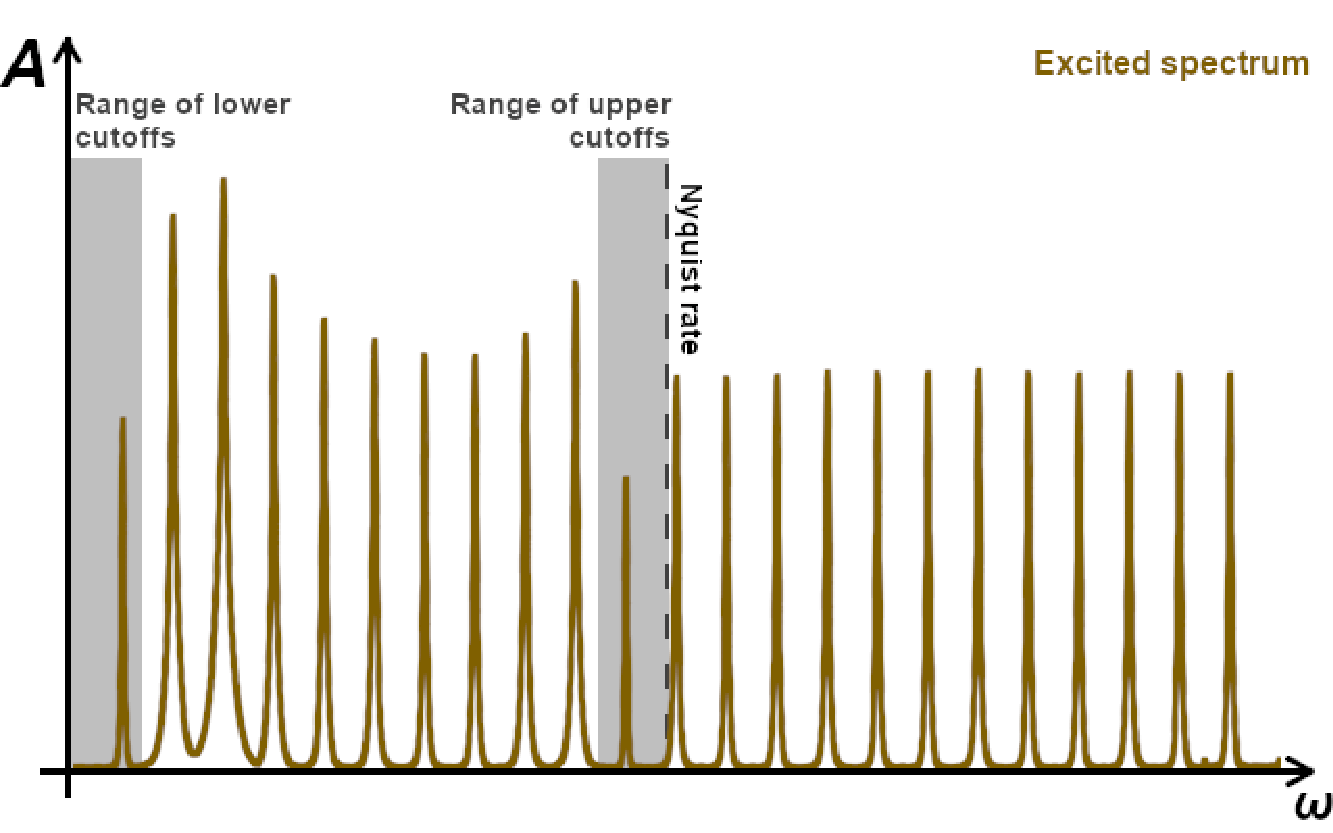}}
	\vspace{-0.8em}
  {(b)}
  \end{minipage}
  \begin{minipage}{.33\textwidth}
	\centering
  \centerline{\includegraphics[width=0.7\columnwidth]{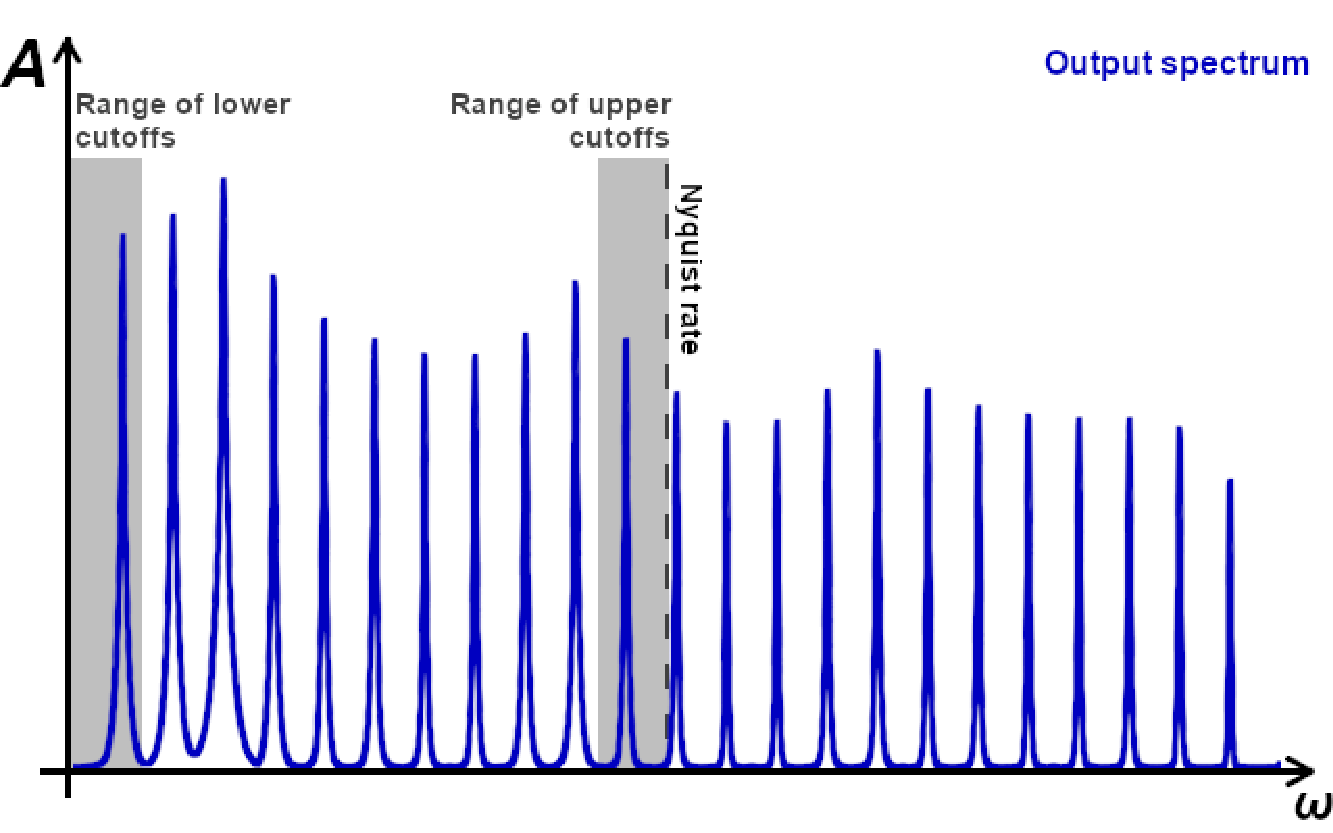}}
	\vspace{-0.8em}
  {(c)}
  \end{minipage}
  \caption{Illustrative (a) full-band/bandlimited spectrum, (b) excited spectrum (per our method) and (c) final output spectrum.}
  \label{fig:bwe}
  	\vspace{-1.0em}
\end{figure*}

\section{Introduction}
\label{sec:intro}

\begin{figure}[htb]

\begin{minipage}[b]{1.0\columnwidth}
  \centering
  \centerline{\includegraphics[width=0.9\columnwidth]{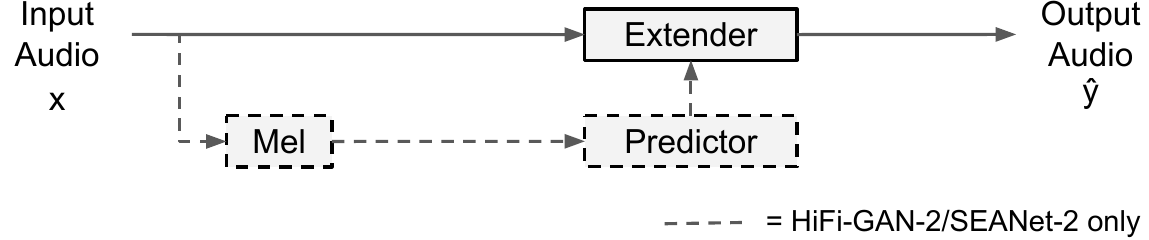}}
%  \vspace{2.0cm}
  \centerline{(a)}\medskip
\end{minipage}
\hfill
\begin{minipage}[b]{1.0\columnwidth}
  \centering
  \centerline{\includegraphics[width=0.9\columnwidth]{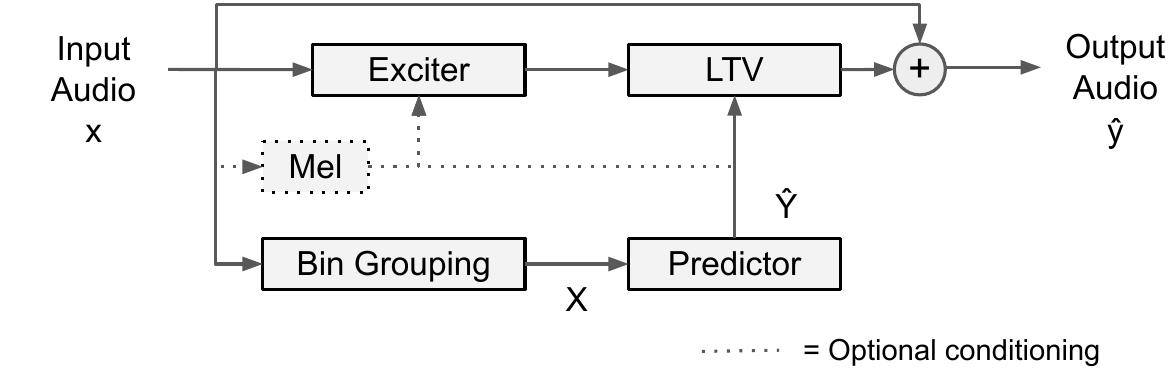}}
%  \vspace{2.0cm}
  \centerline{(b)}\medskip
\end{minipage}
\hfill
\begin{minipage}[b]{1.0\columnwidth}
  \centering
  \centerline{\includegraphics[width=0.9\columnwidth]{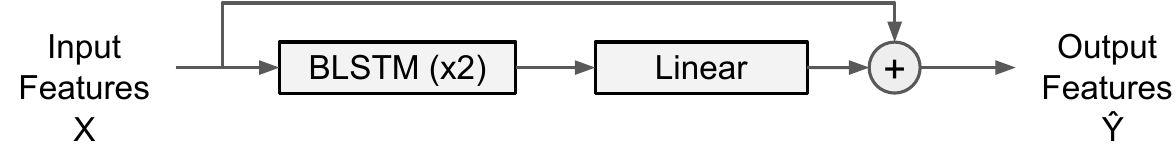}}
%  \vspace{2.0cm}
  \centerline{(c)}\medskip
\end{minipage}
\vspace{-1.0cm}
\caption{(a) Extender, (b) exciter/LTV, (c) predictor models.}
\label{fig:systems}

\vspace{-0.5cm}
\end{figure}

\emph{Bandwidth extension (BWE)} aims at estimating missing frequency components of a signal whose bandwidth has been reduced~\cite{Abel, Old}. Such a reduction is a common artifact of lossy audio codecs, telecommunication channels, or subpar recording gear. In telephony, it is common to see speech bandwidth reduced to 4~kHz~\cite{Seanet, jax_2003}. An inoptimally placed microphone could also attenuate lower frequencies below the noise floor.

Early approaches to BWE attempted to infer spectral envelopes and/or magnitudes of the upper frequency band from the available lower band~\cite{Abel, Old, jax_2003}. The first waveform-to-waveform deep learning model for BWE was introduced in ~\cite{Kuleshov}, which used a convolutional encoder/decoder network.  The approach was upgraded by adding temporal feature-wise linear modulation (TFiLM) to expand the receptive field of convolutional models~\cite{Film}, or by integrating a parallel frequency-domain network alongside of it in order to combine their advantages~\cite{TimeFrequency}.  Moreover, generative adversarial networks (GAN) have been proposed for BWE~\cite{Seanet, Hifigan}, leveraging the MelGAN adversarial training framework~\cite{MelGAN} and using feedforward WaveNet~\cite{Hifigan, OrigHifigan, WaveNet} or SEANet~\cite{OrigSeanet} architectures as generators.  These methods not only promote good fidelity in the average sense during training, but also the plausibility of system outputs, as measured by a discriminator tuned to discern ground truth and synthesized examples. Said approaches combine time-domain, time-frequency domain, and adversarial losses, along with a deep feature loss defined on discriminator feature maps to regulate adversarial training. Recent advances have leveraged diffusion-based generative models \cite{richter_speech_2023}, offering unprecedented quality, and neural codec language model-based approaches \cite{wang_speechx_2023, yang_uniaudio_2023}, providing additional control but at a significant computational cost. We highlight the continued importance of GANs, favored for their compatibility with lightweight architectures ideal for on-edge deployment. Our work systematically explores DSP-informed BWE, positioning it as a complementary approach with broad applicability across various systems.

In this work, we model BWE explicitly as an excitation of the source signal which generates new harmonics, followed by a linear time-varying (LTV) filtering operation which shapes this excitation, all within an end-to-end deep learning context.  We decompose BWE models into exciter and filtering modules, such that the exciter module is primarily tasked with making the signal wideband, whereas the filter module is responsible for sculpting it to achieve the desired spectral profile.  This is a natural formulation for BWE, per the aforementioned approaches~\cite{Abel, Old}; however, to the best of our knowledge, this serial exciter/filter formulation has not yet been used directly in an end-to-end deep learning framework.  Motivated by~\cite{Ddsp}, we demonstrate how the inductive bias provided by the added LTV filter improves system fidelity, and/or allows us to use lighter exciter subnetworks.  Our methodology can be extended to any existing black-box generator candidate and training setup.  Here, we consider the HiFi-GAN and SEANet models as candidate exciters, use a recurrent neural network (RNN) feature predictor~\cite{Hifigan2} to drive the LTV filter, and train models adversarially.

This paper is organized as follows:  Section~\ref{sec:related} outlines related work.  Section~\ref{sec:methods} introduces proposed methods.  Section~\ref{sec:results} illustrates experimental results. Section~\ref{sec:conclusions} draws conclusions.

\section{Related work}
\label{sec:related}

Figure~\ref{fig:bwe} illustrates the BWE task.  Frequencies from a wideband speech signal $y$ are removed during lossy transmission, leaving a signal $x$ with a narrow passband of frequencies intact.  Existing approaches directly attempt to infer the extended result $\hat{y}$, while we consider an intermediate product, as in Figure~\ref{fig:bwe}b, in our dual-stage strategy.  HiFi-GAN/HiFi-GAN-2 use a feedforward variant of WaveNet (temporal convolutional network~\cite{Steinmetz}), while SEANet uses a convolutional encoder/decoder model (1D U-Net~\cite{Unet}).  Models operate on time-domain signals and are trained in an end-to-end fashion using negative log-likelihood and adversarial losses (with discriminator architecture and training setup in~\cite{MelGAN}). HiFi-GAN generally outperforms SEANet, but at a much higher computational cost.  SEANet is lightweight due to its strided convolutions, making it more amenable for real-time use.     

\subsection{HiFi-GAN}

The generator in HiFi-GAN is a feedforward WaveNet~\cite{Hifigan, OrigHifigan, WaveNet} using 3 stacks of 8 dilation layers with a dilation factor of 2, 128 channels, and a kernel size of 3.  Skip channels from all layers are summed, and a postnet transforms the result down to a single channel.  During training, a generator loss compares inferred and ground-truth audio samples $\mathcal{L}_{BWE, G}(y, \hat{y})$, which encompasses the entirety of negative log-likelihood and adversarial loss terms used to train the extender in~\cite{Hifigan}.  The discriminator is trained according to ~\cite{Hifigan, MelGAN}.  Unlike the implementation in~\cite{Hifigan}, we impose the temporal convolutional network to learn a residual signal $\hat{y} - x$, and add its output to the input in order to yield the system output.

\subsection{HiFi-GAN-2}

In~\cite{Hifigan2}, the HiFi-GAN generator was extended to speech denoising and dereverberation.  An RNN acoustic feature prediction subnetwork was added, which infers acoustic features of the enhanced audio---it was particularly beneficial for dereverberation.  The output of the feature predictor is used to locally condition the feedforward WaveNet, resembling the TFiLM approach in~\cite{Film}.  As illustrated in Figure~\ref{fig:systems}a, the \emph{generator} network is now comprised of an \emph{extender} network, extending the bandwidth of its input, and a \emph{predictor} network whose output adapts the behavior of the extender network.  The feature prediction subnetwork efficiently increases the receptive field of the generator since it operates at a frame rate that is orders of magnitude slower than the audio sample rate.

We slightly change the predictor model in~\cite{Hifigan2} to make it more lightweight and learn more efficiently, as shown in Figure~\ref{fig:systems}c.  Our architecture consists of two bidirectional LSTM layers with 128 units, followed by a linear projection down to the feature dimension.  Furthermore, we constrain the input and output features of the model to be of the same type and therefore have the same number of channels.  This allows us to use a residual connection, so that the trainable components of the predictor learn to infer acoustic features for spectral components that are outside of the input passband.  The feature prediction network operates on 80-band log mel spectrograms generated with a 2048-point short-time Fourier transform (STFT) at a frame rate of 93.75~Hz (512-sample hop size at 48~kHz).  Unlike~\cite{Hifigan2}, we train the acoustic feature prediction objective jointly with the BWE task.  Given input acoustic features $\mathbf{X}$, ground truth acoustic features $\mathbf{Y}$, and inferred acoustic features $\mathbf{\hat{Y}} = P(\mathbf{X})$ (where $P$ is the predictor network), the generator loss now becomes
\begin{equation} \label{eqn:mse}
\mathcal{L}_G(y, \hat{y}, \mathbf{Y}, \mathbf{\hat{Y}}) = \mathcal{L}_{G,BWE}(y, \hat{y}) + \mathbf{E}[|\mathbf{Y}-\mathbf{\hat{Y}}|].
\end{equation}

\subsection{SEANet}
In~\cite{Seanet}, a 1D convolutional encoder/decoder generator architecture was proposed for the 8~kHz to 16~kHz sampling rate BWE task.  It consists of pre/post-processing convolutional layers and bottlenecks (shown in green in Figure~\ref{fig:seanet}), as well as complementary encoder/decoder blocks with skip connections (depicted in blue in Figure~\ref{fig:seanet}).  Each encoder block consists of three residual units, where each residual unit contains dilated convolutions followed by a strided convolution which effectively downsamples signals in time.  Each decoder block complements its respective encoder block with transposed convolutions of matching stride to perform signal upsampling, and residual units with matching channel count.  For further details, see~\cite{Seanet, OrigSeanet}.  We change channel and block counts in~\cite{Seanet} to better accommodate the 6x upsampling factor case (see Figure~\ref{fig:seanet} for details).  An additional block is added to the middle of encoder/decoder stacks with a stride of 4.  The resulting [2, 2, 4, 8, 8] stride pattern extends the receptive field of the architecture to 1024 samples (as compared to the 256 samples in~\cite{Seanet}).  Lastly, non-causal convolutions are used as in~\cite{OrigSeanet}, as opposed to the streaming convolutions in~\cite{Seanet}.

\section{Proposed method}

\label{sec:methods}

\subsection{HiFi-GAN-2}
We propose HiFi-GAN-2 as a baseline BWE model to compare against our proposed exciter/filter method.  To our best knowledge, this is the first work to use HiFi-GAN-2 for BWE.

\subsection{Locally-conditioned SEANet (SEANet-2)}

\begin{figure}[t]
\begin{minipage}[b]{1.0\linewidth}
  \centering
  \centerline{\includegraphics[width=0.88\columnwidth]{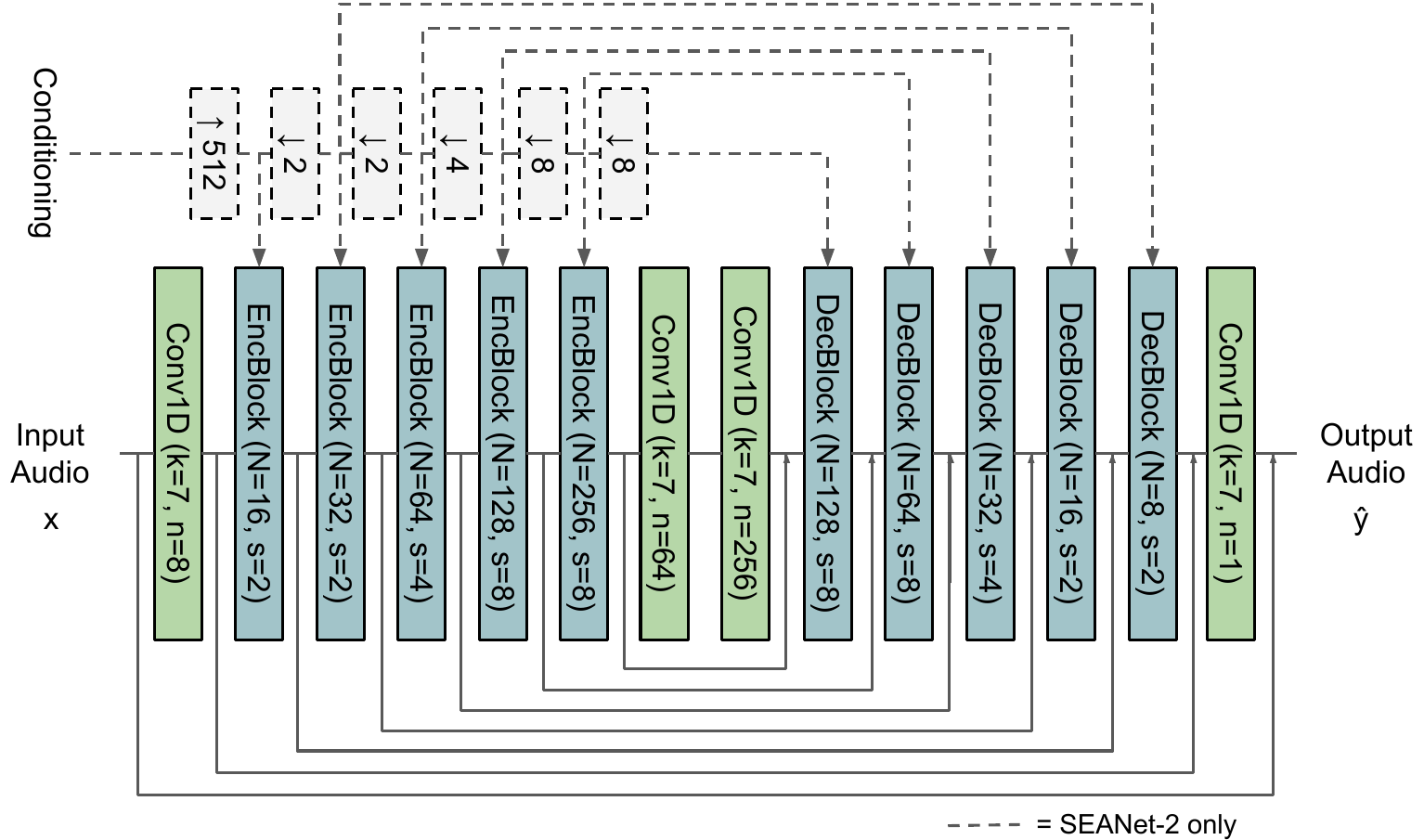}}
\end{minipage}
\vspace{-0.75cm}
\caption{Block diagram for SEANet/SEANet-2 extenders.}
\label{fig:seanet}
\vspace{-0.6cm}
\end{figure}

For completeness, we consider a ``HiFi-GAN-2 equivalent'' SEANet architecture.  While the HiFi-GAN extender lends itself to receive local conditioning information~\cite{WaveNet}, care is needed to extend this to SEANet due to its use of strided convolutions.  When SEANet is paired with a feature predictor network, as in HiFi-GAN-2, we call it ``SEANet-2.'' Figure~\ref{fig:seanet} illustrates the SEANet architecture and our adaptation to include local conditioning.  We begin by upsampling frame-level conditioning information (i.e., acoustic feature predictions, as in HiFi-GAN-2) to audio rate via linear interpolation.  Next, we downsample the upsampled conditioning signal by all the downsampling factors observed in the various SEANet encoder and decoder blocks.  We feed the respective downsampled signal to each encoder and decoder block, omitting conditioning information in bottleneck and outermost convolutional layers.  The conditioning information at each residual unit of each block is transformed via a $1 \times 1$ convolutional layer to match the channel count of the transformed input.  We add this to the transformed input prior to its ELU activation.

\subsection{Serial exciter/filter paradigm}
\label{ssec:paradigm}

As illustrated in Figure~\ref{fig:systems}b, we split the \emph{extender} network into an \emph{exciter} network, broadening the bandwidth of its inputs (per Figure~\ref{fig:bwe}b), and an \emph{LTV filter} which shapes the excited signal spectrum under an end-to-end framework.  This eases the burden for the network to learn the BWE task, by inducing the bias that BWE can be viewed as creating a wideband source excitation and filtering it~\cite{Abel}.  To model the exciter, we can use most any neural network. When using HiFi-GAN or SEANet-style exciters, we refer to our approaches as HiFi-GAN-2 + LTV and SEANet-2 + LTV respectively.  We use a zero-phase filtering methodology for the LTV filter using a differentiable 2048-point STFT with 512-sample hop size, which is driven by the outputs of the feature predictor with matching specifications. %This serial exciter/filter approach used here resembles the vocoder in~\cite{DiffWorld}, except that the order of its sub-networks are reversed.
As indicated by the dotted lines in Figure~\ref{fig:systems}b, the exciter could have been conditioned using the outputs from the predictor model.  However, here we condition it on the mel spectrogram of the input to maintain that the exciter subnetwork remains exactly the same size as its corresponding extender network used for comparison.

To minimize the number of acoustic features inferred by the feature predictor subnetwork, we use a compressed spectral representation that allows us to soundly decompress it to full frequency resolution in a differentiable fashion via a pseudoinverse. For an input $x$, we can compute its magnitude spectrum $\mathbf{F}_x$, and define a coarse log-magnitude spectrum as
\begin{equation}
\mathbf{X} = \log_{10}(\mathbf{M}\mathbf{F}_x+\epsilon),
\end{equation}
where $\epsilon = 10^{-5}$ and $\mathbf{M}$ is a non-negative frequency bin grouping matrix representing a non-overlapping brickwall filter bank with $K = 64$ bands.  The feature prediction network infers a compressed spectral representation $\mathbf{\hat{Y}}= P(\mathbf{X})$ in the same basis.  We expand $\mathbf{\hat{Y}}$ to full frequency resolution via
\begin{equation}
\label{eqn:world_mels_decompress}
\mathbf{F}_{\hat{y}} = \mathbf{M}^\dagger(10^\mathbf{\mathbf{\hat{Y}}}-\epsilon),
\end{equation}
where $\mathbf{M^{\dagger}}$ is the pseudoinverse of $\mathbf{M}$, and use $\mathbf{F}_{\hat{y}}$ as the response for the LTV filter.  The non-overlapping design of $\mathbf{M}$ ensures that $\mathbf{M^{\dagger}}$, and therefore $\mathbf{F}_{\hat{y}}$,  is also non-negative.

Due to the residual connection in Figure~\ref{fig:systems}b, the exciter/filter combination infers the residual signal whose frequency components are outside of the input passband. Therefore, we only apply the acoustic feature loss term to spectral bins at or above the upper frequency cutoff as
\begin{equation} \label{eqn:mse2}
\mathcal{L}_G(y, \hat{y}, \mathbf{Y}, \mathbf{\hat{Y}}) = \mathcal{L}_{G,BWE}(y, \hat{y}) + \mathbf{E}[|\mathbf{Y}_{k:}-\mathbf{\hat{Y}}_{k:}|],
\end{equation}
where $k=11$ for a 6x upsampling factor and $K=64$.  The acoustic feature loss not only aids learning, but also implicitly influences the exciter to produce flat spectra in the upper frequency band (as illustrated in Figure~\ref{fig:bwe}b).  This is to say that if $\mathbf{\hat{Y}}$ is close to the true $\mathbf{Y}$, then the upper frequency spectrum of the exciter output must be flat in order for $\hat{y}$ to resemble $y$.

\section{EXPERIMENTAL RESULTS}
\label{sec:results}

To assess the performance of our proposed methods, we train the following models for the task of 8~kHz to 48~kHz BWE using the VCTK dataset~\cite{vctk}:  (1) HiFi-GAN, (2) HiFi-GAN-2, (3) HiFi-GAN-2 + LTV, (4) SEANet, (5) SEANet-2, (6) SEANet-2 + LTV.  The models were trained for a million steps using Adam optimizers, learning rates of $10^{-4}$, and a sequence length of 65536 samples.  In order to train adversarially, we use the discriminator architecture and adversarial loss terms (hinge/deep feature losses) defined in~\cite{MelGAN}.  We use 90\% of the dataset for training and save the remainder for validation.  We use a batch size of 4 for SEANet-based methods, and must opt for a batch size of 2 for the HiFi-GAN-based methods due to GPU memory constraints. During training, we randomly generate bandlimited examples from wideband audio using brickwall filters defined in the frequency domain with randomly varying lower and upper cutoff frequencies, which improves system robustness~\cite{Seanet} (per the gray bands in Figure~\ref{fig:bwe}).  Lower cutoff frequencies varied between 0 and 500~Hz, whereas upper cutoff frequencies varied between 3.5~kHz and 4~kHz.  For subjective listening and supplemental figures, please visit \href{https://bwe-ltv.netlify.app}{\nolinkurl{https://bwe-ltv.netlify.app}}.

We compute average L1 log mel spectrogram errors, Short-Time Objective Intelligibility (STOI) measures \cite{stoi}, and Contrastive Language-Audio Pretraining (CLAP) scores \cite{wu_large-scale_2023} (using their default pretrained model) between BWE examples and their ground truth targets. We also measure deep feature losses across \emph{every} discriminator that we have trained, as their feature maps should distinguish perceptual cues for BWE.  Doing so across all discriminators ensures that there is no bias towards a matched generator/discriminator pair.  We performed a MUltiple Stimuli with Hidden Reference and Anchor (MUSHRA)~\cite{MUSHRA} test to evaluate models qualitatively.  We prepared 8 trials and asked participants to rank how accurately each test item sounded like its corresponding wideband reference.  To prevent participant fatigue from too many test items, we considered a subset of models, focusing to test into the DSP-informed nature of our exciter/LTV approach relative to prior art.  We opted for HiFi-GAN over the SEANet family of models since HiFi-GAN/HiFi-GAN-2~\cite{Hifigan, Hifigan2} had been established independently from this work.  Trials included a hidden reference, its narrow-band counterpart as an anchor (input to BWE models), and the outputs of HiFi-GAN, HiFi-GAN-2, and HiFi-GAN-2 + LTV. Participants within our organization with critical listening skills conducted tests with their own choice of headphones.
\begin{table}[htb]
\vspace{-1.0em}
\centering
  \begin{tabular}{lccccc}
    \toprule
     Model & Mel & STOI & CLAP & RTF \\ \midrule
     (1) HiFi-GAN & 0.108 & 0.954 & 0.888 & 1.834 \\
     (2) HiFi-GAN-2 & 0.106 & 0.959 & 0.869 & 2.429 \\
     \textbf{(3) HiFi-GAN-2 + LTV} & \textbf{\underline{0.092}} & \textbf{0.968} & \textbf{0.920} & 2.431  \\ \midrule
     (4) SEANet & 0.366 & 0.931 & 0.844 & \textbf{\underline{0.061}} \\
     (5) SEANet-2 & 0.120 & 0.939 & 0.878 & 0.197  \\
     \textbf{(6) SEANet-2 + LTV} & \textbf{0.099} & \textbf{\underline{0.979}} & \textbf{\underline{0.935}} & \textbf{0.216}  \\ \bottomrule
  \end{tabular}
 \caption{Quantitative fidelity metrics and average real-time factors measured on an Intel Xeon 3.1~GHz CPU.}
 \label{tab:quant1}
\vspace{-0.0em}
\end{table}
\begin{table}[htb]
\vspace{-1.0em}
\centering
  \begin{tabular}{ccccccc}
    \toprule
     Model & (1) & (2) & (3) & (4) & (5) & (6) \\ \midrule
     (1) &          10.72 & 10.03 & 13.17 & 7.07 & 3.96 & 4.66 \\
     (2) & \textbf{10.49} & 10.01 & 12.99 & \textbf{6.71} & 4.00 & 4.63 \\
     \textbf{(3)} & 10.56 & \textbf{9.75} & \textbf{12.95} & 6.99 & \textbf{3.85} & \textbf{4.54}   \\ \midrule
     (4) & 11.37 & 10.39 & 14.34 & 7.08 & 4.07 & 4.79 \\
     (5) & 11.34 & 10.84 & 14.22 & 7.18 & 4.31 & 4.93 \\ 
     \textbf{(6)} & \textbf{\underline{10.11}} & \textbf{\underline{9.40}} & \textbf{\underline{12.18}} & \textbf{\underline{6.62}} & \textbf{\underline{3.68}} & \textbf{\underline{4.35}} \\ \bottomrule
  \end{tabular}
 \caption{Feature losses measured across each discriminator.}
 \label{tab:quant2}
 \vspace{-1.0em}
\end{table}
\begin{table}[htb]
\vspace{-1.0em}
\centering
  \begin{tabular}{lccc}
    \toprule
     Model & MUSHRA \\ \midrule
     Reference (full-band) & 90.00 \\
     Anchor (narrow-band) & 27.25 \\ \midrule
     (1) HiFi-GAN & 75.93 \\
     (2) HiFi-GAN-2 & 74.29 \\
     \textbf{(3) HiFi-GAN-2 + LTV} & \textbf{78.42}  \\ \bottomrule
  \end{tabular}
 \caption{Results of our subjective evaluation.}
 \label{tab:qual}
\vspace{-1.0em}
\end{table}

Our results are summarized in Tables~\ref{tab:quant1}-\ref{tab:qual}. Across all fidelity metrics, one of our proposed models using an exciter/LTV filter approach is the top performer.  HiFi-GAN-2 + LTV outperformed all approaches when evaluating fidelity using the mel reconstruction error, with SEANet-2 + LTV as a close second.  SEANet-2 + LTV was the top performer when assessing fidelity based on the deep feature loss metric defined across all discriminators.  We also observe the poor performance of SEANet when used directly for the 6x upsampling task, and how frame-level local conditioning improves efficacy.  With an interest in deploying models locally on consumer machines, average real-time factors (RTFs) show how the SEANet-based approaches run several times faster than real-time on a CPU.  Lastly, our MUSHRA test confirms the superior performance of our exciter/LTV method.

\section{CONCLUSIONS}
\label{sec:conclusions}
We proposed an end-to-end exciter/filter approach to BWE.  We showed how the inductive bias imparted by the proposed method can improve upon existing approaches by both quantitative and qualitative means. In the future, we would like to make improvements to the exciter/filter paradigm, for example, to incentivize excitations to follow a pink noise profile. As our methodology marks an architectural change, we would like to incorporate it into diffusion-based BWE models~\cite{NuWave2}.%, NuWave2}.

\bibliographystyle{IEEEbib}
\bibliography{refs}

\end{document}